\pdfoutput=1
\documentclass[11pt,twoside,a4paper,cmspaper,final,collab]{cms-tdr}
\def\svnVersion{b019519}\def\svnDate{2025/03/22}\def\cmsCernNoTag{CERN-EP-2024-138}\def\cmsCernDate{\today}\def\cmsMessage{Published in Physical Review Letters as \href{http://dx.doi.org/10.1103/PhysRevLett.134.111903}{\doi{10.1103/PhysRevLett.134.111903}.}}
\begin{document}\cmsNoteHeader{HIN-22-001}

\newlength\cmsTabSkip\setlength{\cmsTabSkip}{1ex}
\ifthenelse{\boolean{cms@external}}
{\providecommand{\cmsTable}[1]{#1}}
{\providecommand{\cmsTable}[1]{\resizebox{\textwidth}{!}{#1}}}

\ifthenelse{\boolean{cms@external}}{\providecommand{\suppMaterial}{\cite{suppMatBib}}}{\providecommand{\suppMaterial}{Appendix~\ref{app:suppMat}}}
\newlength{\cmsFigWidth}
\ifthenelse{\boolean{cms@external}}{\setlength\cmsFigWidth{\columnwidth}}{\setlength\cmsFigWidth{0.65\textwidth}}
\ifthenelse{\boolean{cms@external}}{\providecommand{\cmsfigone}{ Fig.~2}}{\providecommand{\cmsfigone}{ Fig.~\ref{fig:css_final}}}
\ifthenelse{\boolean{cms@external}}{\providecommand{\cmstabsupp}{ Table~1}}{\providecommand{\cmstabsupp}{ Table~\ref{table:Systematics_Fonll_sup}}}
\ifthenelse{\boolean{cms@external}}{\providecommand{\cmsfigsupp}{ Fig.~1}}{\providecommand{\cmsfigsupp}{ Fig.~\ref{fig:z2mm_css_Ntrk}}}

\ifthenelse{\boolean{cms@external}}{\providecommand{\cmsLeft}{upper\xspace}}{\providecommand{\cmsLeft}{left\xspace}}
\ifthenelse{\boolean{cms@external}}{\providecommand{\cmsRight}{lower\xspace}}{\providecommand{\cmsRight}{right\xspace}}

\ifthenelse{\boolean{cms@external}}{%
    \newcommand{\refsupplemental}{the supplemental material~\cite{this:supplemental}\xspace}
}{%
	\newcommand{\refsupplemental}{the supplemental material\nocite{this:supplemental}\xspace}
}

\ifthenelse{\boolean{cms@external}}{%
    \newcommand{\whimA}{in the main text\xspace}
}{%
	\newcommand{\whimA}{\xspace}
}

\ifthenelse{\boolean{cms@external}}{%
    \newcommand{\whimB}{of this supplemental material\xspace}
}{%
	\newcommand{\whimB}{\xspace}
}

\newcommand{\dT}{d_{\mathrm{T}}}
\newcommand{\dz}{d_{\mathrm{z}}}
\newcommand{\BuJpsiK}{\ensuremath{\PBp\!\to\! \cPJgy \PKp}\xspace}
\newcommand{\BsJpsiPhi}{\ensuremath{\PBzs \!\to\! \PJgy \Pgf}}

\newcommand{\Bu}{\ensuremath{\PBp}\xspace}
\newcommand{\Bs}{\ensuremath{\PBzs}\xspace}
\newcommand{\Bd}{\ensuremath{\PBz}\xspace}
\newcommand{\Bc}{\ensuremath{\PBpc}\xspace}
\newcommand{\LambdaB}{\ensuremath{\PGLzb}\xspace}

\newcommand{\Jpsik}{\ensuremath{\PJgy \PKp}\xspace}
\newcommand{\Jpsipi}{\ensuremath{\PJgy \Pgpp}\xspace}
\newcommand{\JpsiPhi}{\ensuremath{\PJgy \Pgf}}

\newcommand{\Phikk}{\ensuremath{\Pgf\!\to\! \PKp \PKm}\xspace}
\newcommand{\Kpik}{\ensuremath{\PKst{}^{0}\!\to\! \Pgpp \PKm}\xspace}
\newcommand{\Ka}{\ensuremath{ \PK_{1}}\xspace}

\newcommand{\Jpsi}{\ensuremath{\PJGy}\xspace}

\newcommand{\CSS}{\ensuremath{\rd\sigma/\rd\pt}\xspace}
\newcommand{\pPb}{\ensuremath{\Pp\mathrm{Pb}}\xspace}
\newcommand{\PbPb}{\ensuremath{\mathrm{PbPb}}\xspace}
\newcommand{\RpPb}{\ensuremath{R_{\pPb}}\xspace}
\newcommand{\Rhl}{\ensuremath{R_{\mathrm{HL}}}\xspace}
\newcommand{\Rcp}{\ensuremath{R_{\mathrm{CP}}}\xspace}

\newcommand{\RPP}{\ensuremath{R_{\mathrm{PbPb}}}\xspace}
\newcommand{\RBpPb}{\ensuremath{R^{\PBp}_{\pPb}}\xspace}
\newcommand{\RZpPb}{\ensuremath{R^{\PZ}_{\pPb}}\xspace}
\newcommand{\RBhl}{\ensuremath{R^{\PBp}_{\mathrm{HL}}}\xspace}

\newcommand{\xsZ}{\ensuremath{\sigma^{\PZ}}\xspace}
\newcommand{\xsB}{\ensuremath{\sigma^{\PBp}}\xspace}
\newcommand{\CRprime}{\ensuremath{\xsB/\xsB_{0}}\xspace}
\newcommand{\CRprimeZ}{\ensuremath{\xsZ/\xsZ_{0}}\xspace}

\newcommand{\NtrkZ}{\ensuremath{N_{\text{trk}^0}}\xspace}
\newcommand{\Ntrk}{\ensuremath{N_{\text{trk}}}\xspace}
\newcommand{\Ntrkoff}{\ensuremath{\Ntrk^{\text{offline}}}\xspace}
\newcommand{\Ntrkcorr}{\ensuremath{\Ntrk^{\text{corrected}}}\xspace}
\newcommand{\Rprime}{\ensuremath{\sigma_{\Bu}/\sigma^0_{\Bu}}\xspace}

\newcommand{\Nch}{\ensuremath{N_{\text{ch}}}\xspace}
\newcommand{\Ncoll}{\ensuremath{\langle N_{\text{coll}}\rangle}\xspace}
\newcommand{\GFch}{\ensuremath{\mathrm{GF}_{N_{\text{ch}}}}\xspace}

\cmsNoteHeader{HIN-22-001}

\title{Search for nuclear modifications of \texorpdfstring{\PBp}{B+} meson production in \Pp-Pb collisions at \texorpdfstring{$\sqrtsNN = 8.16\TeV$}{sqrt(sNN) = 8.16 TeV}}

\date{\today}

\abstract{
Nuclear medium effects on \PBp meson production are studied using the
binary-collision scaled cross section ratio between events of different charged-particle multiplicities from proton-lead collisions. 
 Data, collected by the CMS experiment in 2016 at a nucleon-nucleon center-of-mass energy of $\sqrtsNN = 8.16\TeV$, 
corresponding to an integrated luminosity of 175\nbinv, were used. The scaling factors in the ratio are determined using a novel approach based on the $\PZ\to\PGmm\PGmp$ cross sections measured in the same events. The scaled ratio for $\PBp$ is consistent with unity for all event multiplicities, putting stringent constraints on nuclear modification for heavy flavor.}

\hypersetup{
pdfauthor={CMS Collaboration},
pdftitle={Search for nuclear modifications of B+ meson production in pPb collision at sqrt(sNN) = 8.16 TeV},
pdfsubject={CMS},
pdfkeywords={CMS, Heavy-ions, B-physics, multiplicity, nuclear medium}}

\maketitle 
In the presence of extreme density and temperature, quantum chromodynamics (QCD) predicts the transition of nuclear matter into a new state, 
called quark-gluon plasma (QGP)~\cite{Shuryak:1977ut,Collins:1974ky}. 
In this state, quarks and gluons, normally confined within particles such as protons and neutrons, are no longer bound and can move over distances exceeding nucleonic scales. 

Evidence for QGP formation came initially from heavy ion collisions, 
showing  particle correlations consistent with the creation of an almost perfect fluid and differences in particle yields compared to expectations based on scaling proton-proton ({\Pp}{\Pp}) collision results by the number of binary collisions in the larger system~\cite{PHENIX:2004vcz,STAR:2005gfr, Busza:2018rrf}. 
The conditions needed for QGP formation were not expected to exist in collisions of smaller systems. 
However, long-range correlations, similar to those found in large system collisions, 
have now also been observed for a number of smaller systems, including proton-lead (\pPb)~\cite{CMS:2024krd, CMS:2012qk, Aad:2012gla, alice:2012qe, Aaij:2015qcq, ABELEV:2013wsa, Khachatryan:2014jra, Khachatryan:2015waa, ALICE:2013wgn, ATLAS:2018ngv, CMS:2018duw, CMS:2017exb, PHENIX:2018lia, CMS:2023rev} 
 and even {\Pp}{\Pp} collisions~\cite{Khachatryan:2010gv, Aad:2015gqa, Khachatryan:2015lva, Khachatryan:2016txc, CMS:2017kcs, ALICE:2018pal,ALICE:2019oyn, CMS:2023iam}. 
 These small system results have led to the suggestion of the creation of quark-gluon droplets, 
 although alternative mechanisms have also been suggested~\cite{Dusling:2015gta,Nagle:2018nvi,CMS:2020qul}. 
 Further studies are needed to establish the mechanism responsible for the small system observations.

Heavy quarks, such as \PQc and \PQb quarks, are invaluable tools for studying QGP~\cite{Dong:2019byy, Apolinario:2022vzg}. 
 Produced early in the collision, they experience the entire lifetime of the medium, allowing them to probe the media properties throughout its evolution. 
 Their large mass minimizes distortions arising from final-state interactions with other particles. 
 Additionally, well-understood perturbative QCD enables precise calculations of their initial production rates. 
 By studying how the medium affects heavy quarks, for example, in terms of energy loss and fragmentation modification, the properties of the medium can be determined. 
 Evidence for collective effects involving charm and beauty hadrons in small systems has been observed~\cite{CMS:2018duw,ALICE:2019oyn, CMS:2020qul,CMS:2023dse}. However, the rapidity dependence of collective effects for beauty hadrons remains poorly constrained~\cite{PhysRevD.99.052011}.

This Letter reports the measurement of the \PBp meson differential cross sections for different charged-particle multiplicity (as defined later) classes of \pPb collision data at a nucleon-nucleon center-of-mass energy of $\sqrtsNN = 8.16\TeV$, 
recorded by the CMS detector at the CERN LHC in 2016.
The event sample corresponds to an integrated luminosity of 175\nbinv. 
A new observable, \Rhl, is introduced. 
It is defined as the ratio of cross sections for events at high multiplicity to those at low multiplicity,
 scaled by the corresponding number of binary collisions in two multiplicity bins. 
 This observable is used to characterize the nuclear medium effects in the heavy flavor sector.
 The original feature of this work is the experimental derivation of the scaling factors appearing in the \Rhl value using the measured \PZ boson production cross sections from the same events. 
 Conceptually, this new observable is similar to the \Rcp ("central-to-peripheral") ratio previously reported in the literature~\cite{STAR:2003wqp,PHOBOS:2004juu,Rusnakova:2016mnu}.
However, it presents the distinct advantage to allow the characterization of nuclear medium effects to rely entirely on data.
 Biases in production processes are independent of multiplicity bins and cancel out in the double ratio. 
 This allows the double ratio to reveal any medium-induced bias in \PBp events.
 Throughout this Letter, charge-conjugate modes are implied unless explicitly noted. 
 The measurement is based on the exclusive decay channel \BuJpsiK, where the \PJgy meson decays to a pair of oppositely charged muons.
 The measured inclusive cross section is compared to fixed-order next-to-leading logarithmic (FONLL) calculations~\cite{Cacciari_2012,Cacciari_2001,Cacciari_1998}.
 Tabulated results are provided in the HEPData record for this analysis~\cite{hepdata}.

The CMS apparatus~\cite{CMS:2008xjf} is a multipurpose, nearly hermetic detector, designed to trigger
on~\cite{CMS:2020cmk,CMS:2016ngn} and identify electrons, muons, photons, and (charged and neutral)
hadrons~\cite{CMS:2020uim,CMS:2018rym,CMS:2014pgm}. A global ``particle-flow" algorithm~\cite{CMS:2017yfk} aims
to reconstruct all individual particles in an event, combining information provided by the all-silicon inner tracker and by the crystal
electromagnetic and brass-scintillator hadron calorimeters, operating inside a 3.8\unit{T} superconducting solenoid, with data from
the gas-ionization muon detectors embedded in the flux-return yoke outside the solenoid. 
 The reconstructed particles are used to build \PGt leptons, jets, and missing transverse momentum~\cite{CMS:2018jrd,CMS:2016lmd,CMS:2019ctu}.
The forward hadron (HF) calorimeter uses steel as absorber and quartz fibers as the sensitive material. 
The two halves of the HF are located 11.2\unit{m} from the interaction region, one on each end, and together they provide coverage in the pseudorapidity range $3.0 < \abs{\eta} < 5.2$. 
They also serve as luminosity monitors.
Events are filtered using a two-tiered trigger system~\cite{CMS:2016ngn}. The first level (L1), composed of custom hardware processors, uses information from the calorimeters and muon detectors~\cite{CMS:2020cmk}. 
The second level, known as the high-level trigger, consists of a farm of processors running a version of the full event reconstruction software optimized for fast processing and reduces the event rate to around 1\unit{kHz} before data storage~\cite{CMS:2016ngn}.

The events are selected with an L1 trigger algorithm requiring the presence of two muon candidates, with no explicit momentum threshold. 
For the off-line analysis, events must pass a set of selection criteria designed to reject background events (from beam-gas collisions), as described in Ref.~\cite{Khachatryan:2016odn}. 
Events are required to have at least one reconstructed primary interaction vertex (PV), formed by two or more tracks, 
with a distance from the center of the nominal interaction region of less than 25\cm along the beam direction and 0.2\cm in the plane transverse to the beam direction. In order to select inelastic hadronic collisions, the \pPb{} events are also required to have at least one tower in each of the HF detectors with energy deposits of more than 3\GeV.

The muons used to reconstruct the \PJgy meson candidates must satisfy the \emph{soft-muon} identification requirements~\cite{CMS:2018rym}  
 and the following kinematic constraints: 
the transverse momentum $\pt>3.4\GeV$ for $\abs{\eta}<0.3$, 
$\pt>3.3\GeV$ for $0.3<\abs{\eta}<1.1$, 
$\pt >  (5.5 - 2 \abs{\eta}) \GeV$ for $1.1<\abs{\eta}<2.1$,
and $\pt>1.3\GeV$ in the $2.1<\abs{\eta}<2.4$ region. 
The muon candidates are combined in opposite-charge pairs to form the \Jpsi{} candidates, which are retained in the analysis if they have an invariant mass between 2.9 and 3.3\GeV 
and an absolute rapidity ($\abs{y}$) less than 2.4. 
The estimate of the p value of the fit (sometimes referred to as $\chi^2$ probability) of both muons coming from the same vertex must be greater than 1\%. 

\PBp mesons are reconstructed by combining a \PJGy candidate (from two muons) with a single \emph{high-purity} track~\cite{CMS:2014pgm} 
with at least 11 hits (including both strip and pixel hits), 
\pt relative uncertainty less than 10\%, $\abs{\eta}< 2.4$, and $\pt > 1\GeV$. 
The two muons and the track are fit together,
assuming they originate from a common vertex (\PBp meson decay point) and constraining the dimuon invariant mass to the \PJGy meson world-average mass~\cite{ParticleDataGroup:2022}. 
The \PKp mass~\cite{ParticleDataGroup:2022} is assigned to the single track. 
The PV is redefined excluding the tracks of the \PBp meson candidates, then the \PBp meson decay length is computed as the distance between the PV and the \PBp decay vertex. 
The analysis is restricted to \PBp mesons in the phase space window of $3 < \pt < 50\GeV$ and $\abs{y}<1.8$, with a decay length greater than 7 times its uncertainty, 
and a p value of the vertex fit greater than 7\%.
If there is more than one \PBp meson candidate in an event, 
the candidate with the highest p value of the vertex fit is selected.

The selection criteria described above are optimized using Monte Carlo simulated samples, 
 which are also used to evaluate the detection efficiency and the shapes of the invariant mass distributions of background components.
 The simulations include signal events of \BuJpsiK decays and background events from other \PQb hadron decay chains that involve \PJGy mesons in the final state.  
Events are generated with \PYTHIA~8.212~\cite{Pythia}, which handles the production and hadronization steps.  
 The \EVTGEN~1.5 package~\cite{EvtGen} is used to simulate the decay of the \PQb and \PQc hadrons, 
 whereas \PHOTOS~215.5~\cite{PHOTOS2} is used for the final-state radiation modeling.
The impact of the CMS detector on the generated events, 
including the trigger and reconstruction steps, 
is simulated with the \GEANTfour package~\cite{Geant4}, 
using algorithms identical to those used in the processing of data.
Each \PYTHIA event is embedded into a simulated \pPb collision event generated with EPOS LHC~1.99~\cite{Epos2015},
which is tuned to reproduce global event properties,
such as the charged-hadron \pt spectrum and particle multiplicity~\cite{Khachatryan:2016odn}.

The event-by-event charged-particle multiplicity is defined using tracks originating from the PV, 
 which must be of high purity and satisfy the following criteria: 
the impact parameter significance [$d/\sigma(d)$] measured with respect to the PV must be less than 3, both along [$\dz/\sigma(\dz)$],
and transverse [$\dT/\sigma(\dT)$] to the beam direction, and 
the relative \pt{} uncertainty $\sigma(\pt)/\pt$ must be less than 10\%. To ensure high tracking efficiency and to reduce the rate of misreconstructed tracks, 
the requirements $\abs{\eta}<2.4$ and $\pt>0.4\GeV$ must be satisfied.
The quantity $\Nch$ is the corresponding multiplicity corrected for detector and algorithm inefficiencies in the same kinematic region.
 Events are divided into classes based on $\Nch$. 
The fraction of the total sample found in each multiplicity interval and the average number of tracks after accounting for the corrections are listed in Table~\ref{table:Ntracks}. 
The uncertainty in the average value is 2.4\%, coming from tracking efficiency uncertainties~\cite{tracking}.

\begin{table}[ht] 
\centering
\topcaption{Fraction of the full event sample that contains a \PBp meson candidate in each multiplicity class, and corrected multiplicity for charged particles with $\abs{y}<2.4$ and $\pt> 0.4\GeV$. 
	Systematic uncertainties are given for the corrected multiplicities, while statistical uncertainties are negligible.}\label{table:Ntracks} 
\renewcommand{\arraystretch}{1.2}
\begin{scotch}{lcc} 
Multiplicity class 	&  Fraction (\%)  & $\langle \Nch \rangle$  \\ 
\hline 
$2 \leq \Nch<250$  	&  100.0  &  $102 \pm 2$   \\
$2 \leq \Nch<60$  	&  19.5   &  $ 43 \pm 1$   \\
$60\leq \Nch < 85$  	&  21.1   &  $ 73 \pm 2$   \\ 
$85\leq \Nch< 110$  	&  20.4   &  $ 97 \pm 2$   \\ 
$110\leq \Nch< 250$  	&  39.1   &  $149 \pm 4$   \\
\end{scotch} 
\end{table}

The differential cross section for the \Bu{} meson production is defined as
\begin{equation} \label{eq:ratioequation1}
	\frac{\rd\sigma}{\rd\pt} = \frac{1}{2} \frac{1}{\Delta \pt}  \frac{N(\pt)}{\epsilon\,\mathcal{B} \lumi}  ,
\end{equation}
where $N(\pt)$ is the measured yield of the \Bu{} meson in a given \pt bin, obtained from a maximum likelihood fit;
$\Delta \pt$ is the corresponding bin width; 
 and $\mathcal{B}$ is the product of the world-average values
for the \BuJpsiK{} and $\PJgy \to \PGmp\PGmm$ branching fractions
$(1.026 \pm 0.031)\times10^{-3}$ and  $(5.961 \pm 0.033)\times10^{-2}$~\cite{ParticleDataGroup:2022}, respectively.
The factor of 2 in the denominator reflects the choice used to quote the
cross section for a single charge (taken to be the \Bu), whereas $N(\pt)$ includes both charge states.
Also, $\epsilon$ is the total efficiency to reconstruct \Bu{} meson candidates, calculated for each bin, and \lumi is the integrated luminosity.

\begin{figure}[t!]
\centering
\includegraphics[width=0.49\textwidth]{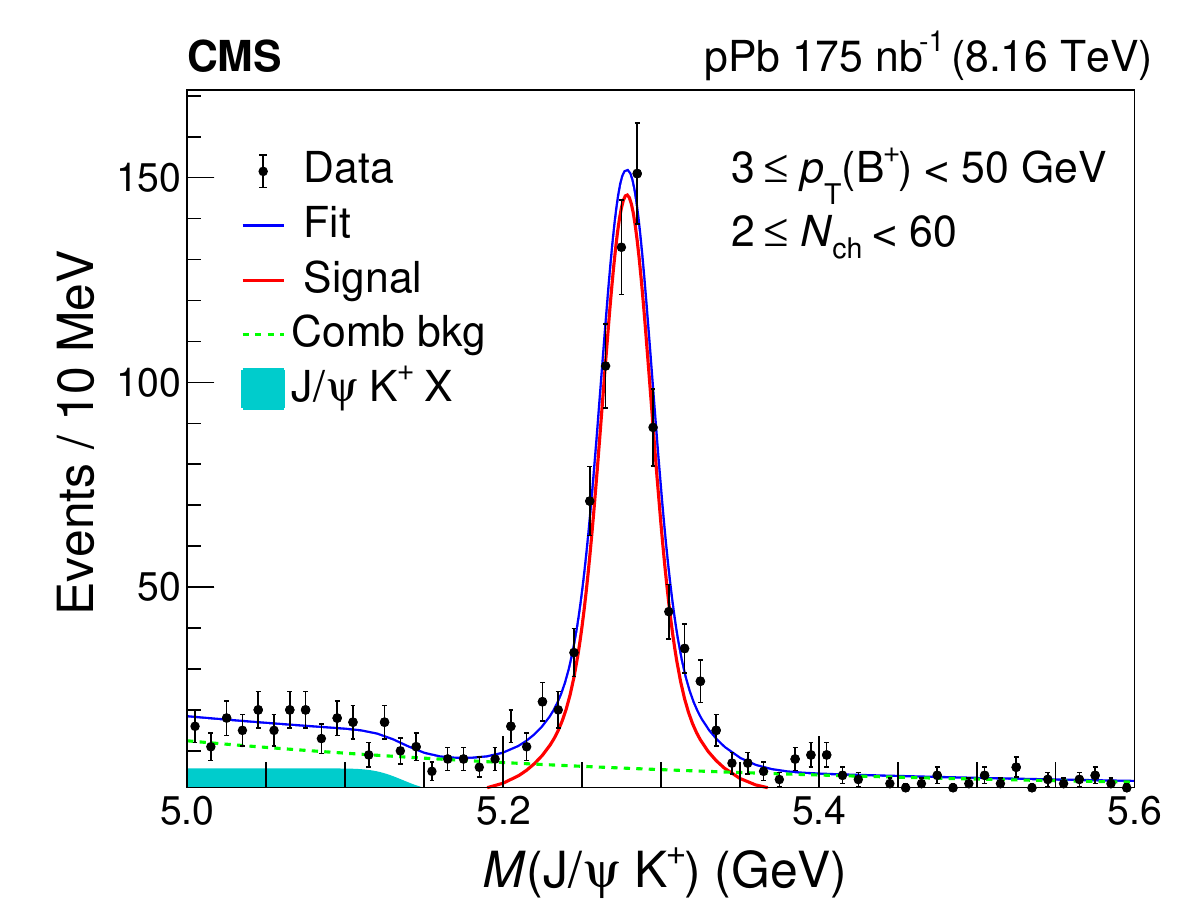}
\includegraphics[width=0.49\textwidth]{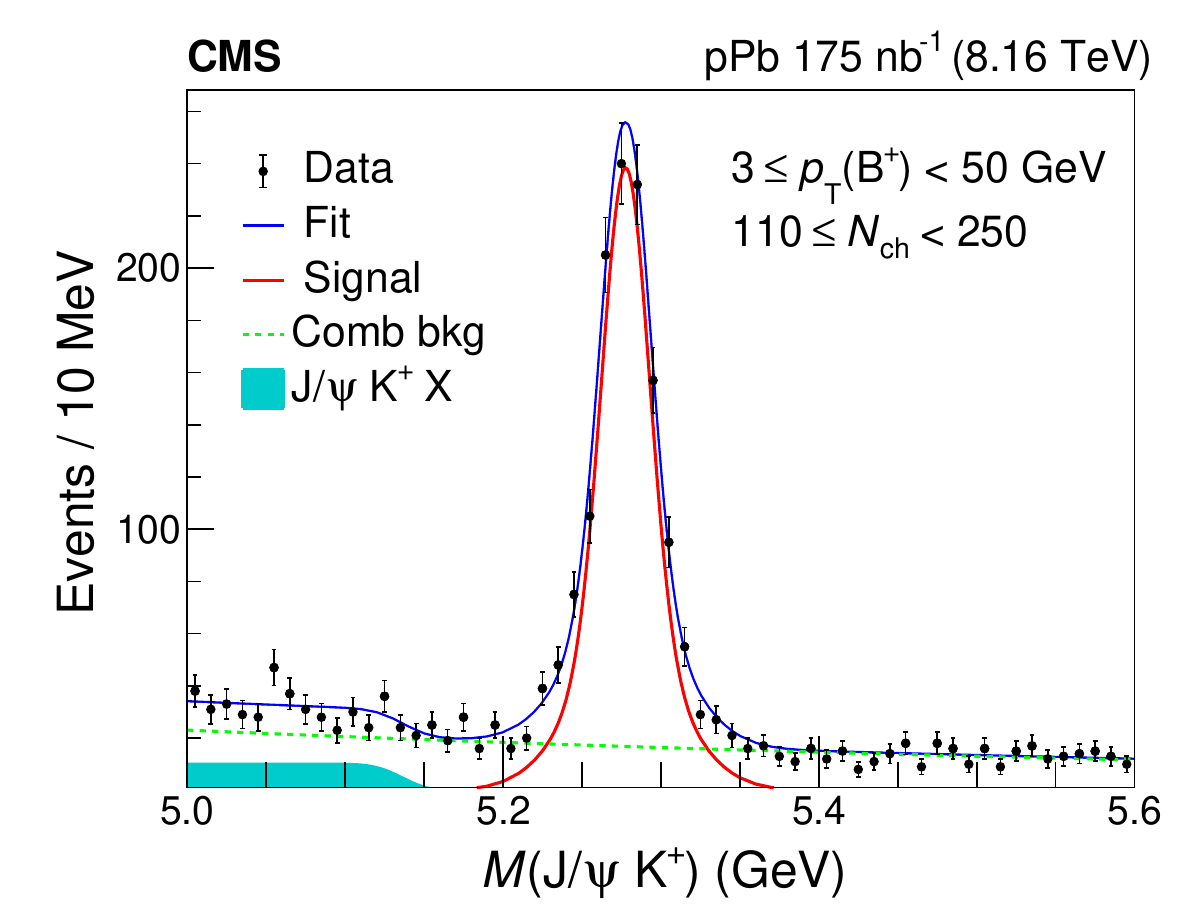}
\caption{Reconstructed \Jpsik invariant mass distributions for \Bu{} meson candidates with $3<\pt<50\GeV$, 
 in events with $ 2 < \Nch  <60 $ (\cmsLeft) or  $ 110 < \Nch  <250 $ (\cmsRight). The result of the fit described in the text is also shown.}
\label{fig:mass-distributions}
\end{figure}

The reconstructed \Bu meson yields are obtained by performing an unbinned maximum likelihood fit to the \Jpsik invariant mass distribution. 
Results are obtained for six \pt bins (see \cmstabsupp{} in \refsupplemental for the binning definition) of four \Nch classes,
with ranges from 2--60 to 110--250, defined so as to keep a similar total event fraction in each class as shown in Table~\ref{table:Ntracks}. In total 24 fits are performed.
Figure~\ref{fig:mass-distributions} shows the \Jpsik invariant mass distributions in the inclusive \pt range 3--50\GeV for two \Nch classes.

On the basis of studies with simulated samples, the signal peak (red curve) is modeled with the sum of two Gaussian probability density functions with 
a common mean and independent widths. 
The common mean is a free parameter in the fit, 
while the widths and their relative fraction are estimated from simulation and fixed in the fit.
 The underlying combinatorial background is fit with an exponential distribution (green dashed curve). 
 The \Jpsik sample includes a background contribution from events 
where the \Bu meson decays to $\PJGy\PKp X$, where $X$ is not reconstructed and represents a \Pgpz or charged light hadrons.
This contribution is described by an error function (cyan area), where the two shape parameters are fixed from simulation.

Detection efficiencies, $\epsilon$, 
are needed to convert the signal event yields, 
obtained from the fits illustrated in Fig.~\ref{fig:mass-distributions}, to cross sections.
 They are evaluated using the simulated event samples, reflecting the detector acceptance, as well as the trigger and reconstruction steps.
 In each multiplicity and \pt bin, the efficiency is calculated as the ratio of events that fire the trigger, are successfully reconstructed by the detector simulation, 
 and satisfy our selection criteria, to the total number of generated \Bu mesons within the specified phase space region $3 < \pt(\PBp) < 50\GeV$ and $\abs{y(\PBp)}<1.8$. 
 It ranges from 0.1\% for $\pt(\PBp)\approx 5\GeV$ to 45\% for $30 < \pt(\PBp) < 50\GeV$.
 In addition to efficiencies derived from simulation, corrections for discrepancies between data and simulated events are applied.
 These corrections were derived from \PJGy meson events in \pPb collisions,using a tag-and-probe technique as described in Ref.~\cite{CMS:2010nis}.

Systematic uncertainties affect the determination of the signal yields and the evaluation of the efficiency [see Eq.~(\ref{eq:ratioequation1})].
The uncertainties affecting the signal yields are evaluated by 
repeating the fits of the mass distributions with alternative models and computing 
the difference with respect to the nominal results.
Three main variations of the fit model are independently considered. 
First, the signal model uncertainty is assessed by leaving free the mean and widths of the double-Gaussian function in the fit. 
Additionally, alternative models, like Johnson $S_{\mathrm{U}}$ distribution~\cite{Johnson:1949zj} and Student's t distribution~\cite{Studentt}, are considered. 
The maximum deviation in measured yields from their nominal value is regarded as a systematic uncertainty.
Second, the combinatorial background is represented by a Chebyshev polynomial instead of an exponential function.
Finally, the uncertainty arising from the description of the partially reconstructed background is quantified by allowing the error function parameters to vary freely in the fit.
The fit procedure is verified to provide unbiased results
for both the central values and the uncertainties,
through a study of 1000 event samples,
randomly generated using the nominal likelihood 
probability distribution functions and fit parameters.
The systematic uncertainties in the signal yields, amounting to up to 4\% in  one of the \pt bins, are dominated by the signal modeling.
The systematic uncertainty in the efficiency ranges from 0.6 to 1.6\% depending on \pt, 
 reflecting the statistical precision in the simulated event samples. 
The uncertainty in the efficiency of the muon trigger, reconstruction, and identification is evaluated using a tag-and-probe technique. 
The data-to-simulation scale factors are employed to determine the nominal efficiency, 
while their associated uncertainties are propagated to the systematic uncertainty by varying coherently the scale factors. 
The difference in track reconstruction efficiency between data and simulation, estimated by the method in Ref.~\cite{tracking},
 results in a 2.4\% uncertainty common to all \pt and \Nch classes.  
Residual differences between \Nch and the  generator-level charged-particle multiplicity distributions are observed in the simulation,  
resulting in a systematic uncertainty in the range 1.3--19.2\%, depending on the  multiplicity class.
In addition, a global systematic uncertainty is assigned to
account for the uncertainties in the integrated luminosity
value (3.5\%~\cite{cmslumi}) and in the \PBp meson branching fraction (3.1\%~\cite{ParticleDataGroup:2022}).

The total systematic uncertainties are computed as 
the sum in quadrature of the individual contributions.
 They range between 2.8 and 14.5\% for the differential cross section measurements, while statistical uncertainties are in the range 3.6--11.0\%.
 All systematic uncertainties are assumed to be uncorrelated between multiplicity classes. 
 
\begin{figure}[t!]
  \centering
  \includegraphics[width=0.49\textwidth]{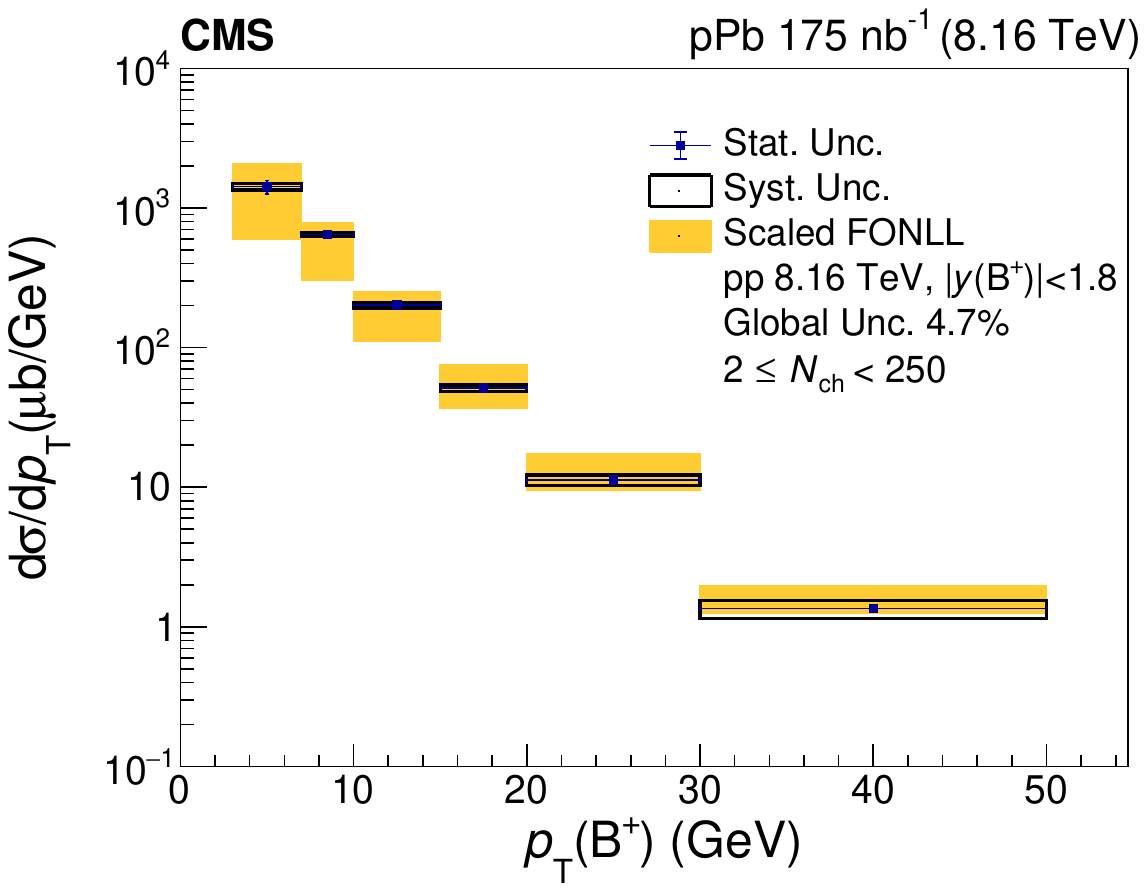}
  \includegraphics[width=0.49\textwidth]{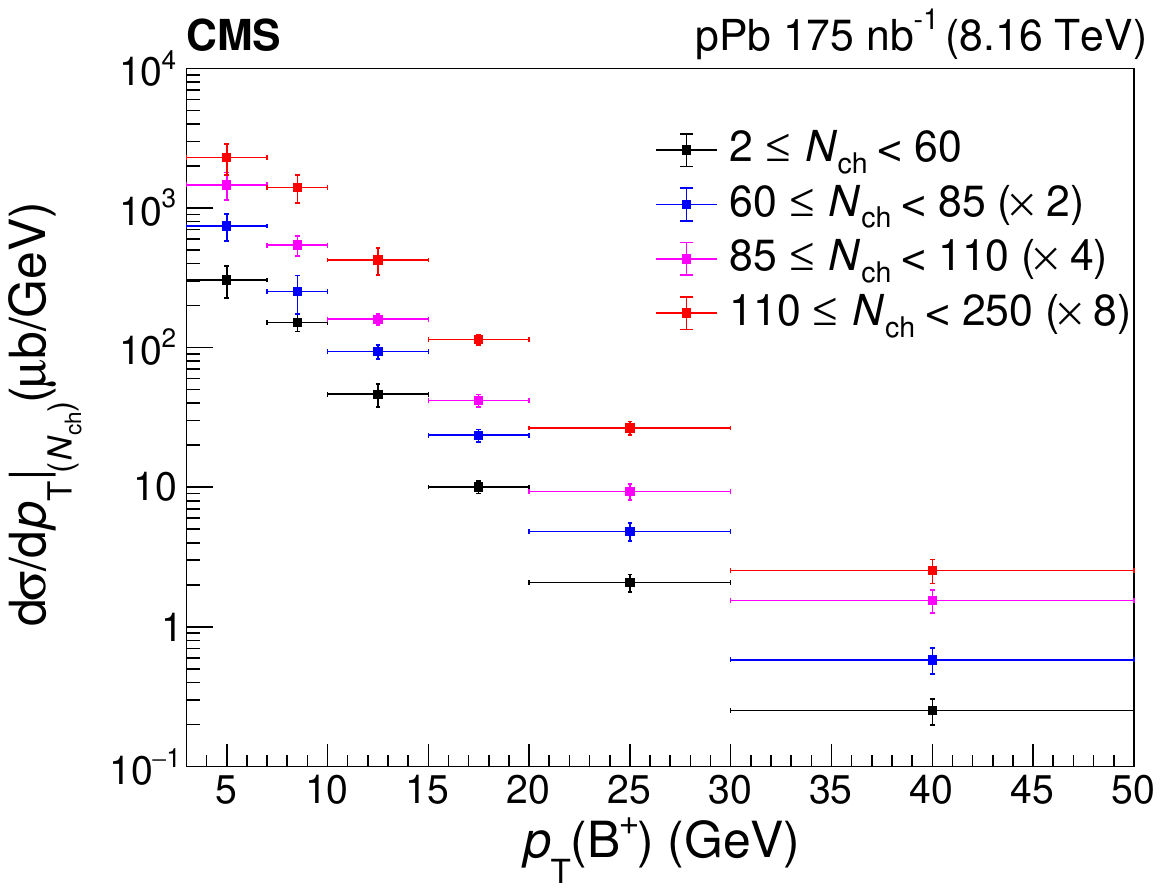}
  \caption{(\cmsLeft) \Bu{} differential cross section in \pt bins.
 Vertical bars (boxes) represent the statistical (systematic) uncertainties.
 The global systematic uncertainty, not included in the data points,
 comprises the uncertainties in the integrated luminosity measurement and the \PBp meson branching fraction.
	Results are compared to FONLL calculations, scaled by A (=208), the mass number of Pb, to account for the number of binary nucleon-nucleon collisions~\cite{Miller:2007ri}. 
 The yellow boxes represent the theoretical uncertainties from FONLL calculations~\cite{Cacciari_2012,Cacciari_2001,Cacciari_1998}. 
 (\cmsRight) \Bu{} differential cross section shown in \pt bins divided into classes of \Nch. 
 For better visibility, data points are scaled by a factor of 2 (blue), 4 (magenta), or 8 (red). 
 Vertical bars represent total uncertainties.
 }
  \label{fig:css_final}
\end{figure}

The differential cross section measurements and its statistical and systematic uncertainties for the phase space region $3 < \pt{}(\PBp) < 50 \GeV$ and $\abs{y(\PBp)} < 1.8$ are displayed in Fig.~\ref{fig:css_final} (\cmsLeft).
 The scaled $\Pp\Pp$ FONLL predictions~\cite{Cacciari_2012,Cacciari_2001,Cacciari_1998} are also shown. 
 The theoretical cross sections are obtained by scaling the FONLL total \PQb quark production by the world-average \Bu{} production fraction of 40.2\%~\cite{ParticleDataGroup:2022}.  
 The predictions are scaled by the mass number of the Pb nucleus ($\mathrm{A}=208$) to consider the number of binary nucleon-nucleon collisions.
 The uncertainties in the FONLL calculations include the effects from the renormalization and factorization scales, the mass of the \PQb quark, 
 and the uncertainties in the proton parton distribution functions (providing the largest contribution)~\cite{Cacciari_2012,Cacciari_2001,Cacciari_1998}. 
 The theoretical predictions are in agreement with the inclusive measurement. 
 Figure~\ref{fig:css_final} (\cmsRight) shows the differential cross section in \pt bins divided into classes of charged-particle multiplicity. 

\begin{figure}[hbt!]
  \centering
  \includegraphics[width=0.49\textwidth]{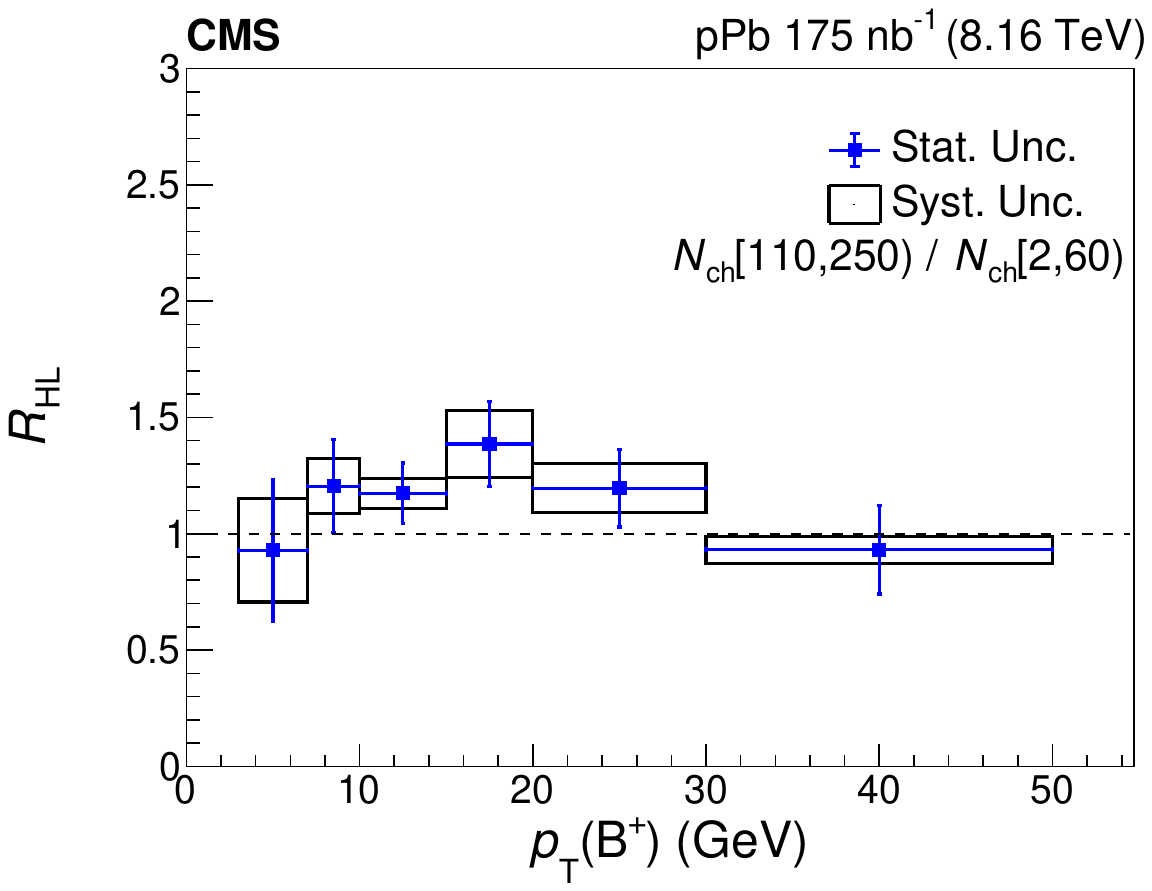}
  \includegraphics[width=0.49\textwidth]{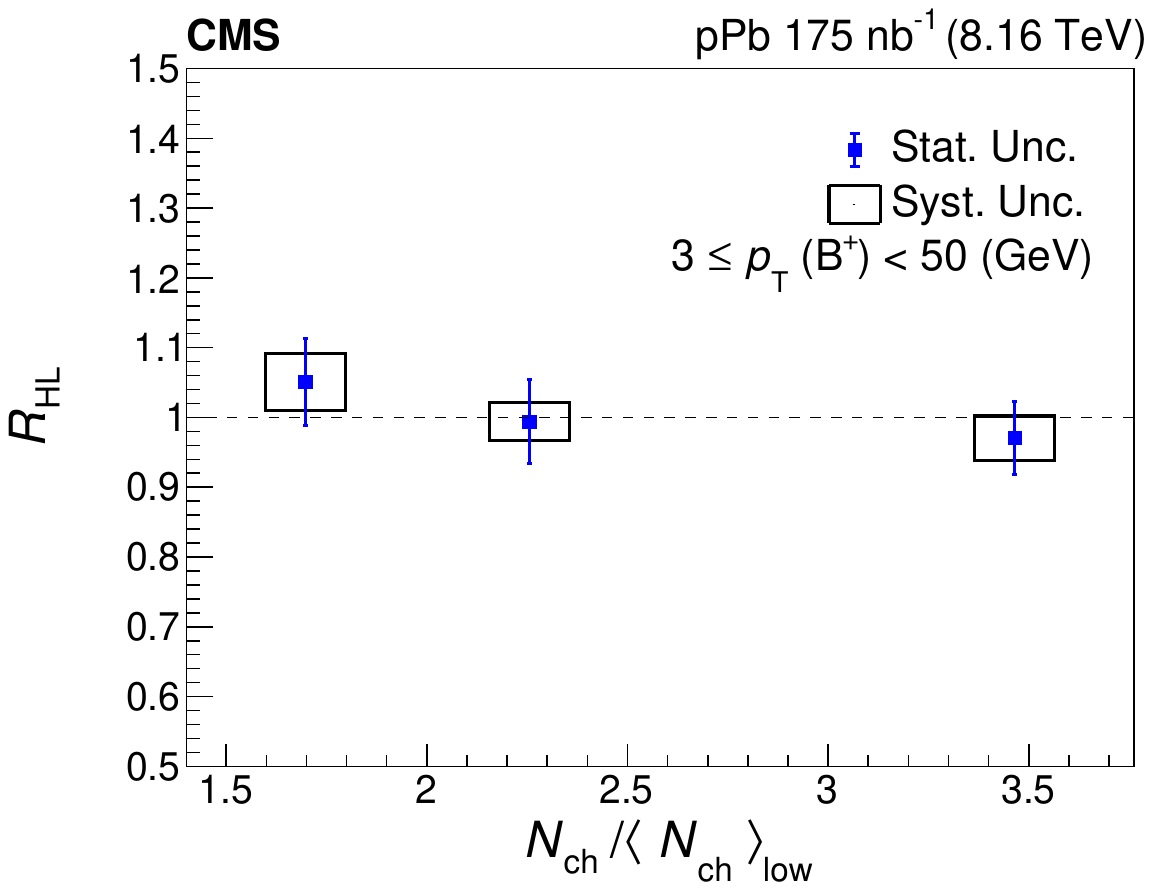}
  \caption{
The $\Rhl$ for \PBp in \pt bins for the highest and lowest multiplicity classes (\cmsLeft) and in the full \pt range and as a function of the multiplicity density (\cmsRight). 
    The error bars correspond to the statistical uncertainty, and the boxes represent the sum in quadrature of systematic uncertainties.
  }
  \label{fig:z2mm_ratios_scaled_gf_GF}
\end{figure}

The effect of the nuclear medium on \PBp meson production is examined through the \Rhl variable conceptually defined as:
\begin{equation}\label{eq:def_rhl}
\Rhl = \frac{\Ncoll|_{\text{low}}}{\Ncoll|_{\text{high}}}\frac{(\CSS)|_{\text{high}}}{(\CSS)|_{\text{low}}} ,
\end{equation}
where, $\Ncoll$ is the average number of incoherent nucleon-nucleon collisions in the events of a certain multiplicity class, 
and \CSS is the differential cross section as a function of the \pt at high (numerator) or low (denominator) multiplicity.
 The production of \PZ bosons is expected to scale with the number of binary collisions and be void of any final-state medium effects;
 thus, the ratio of \PZ boson cross section in the same high- and low-multiplicity event classes provides experimental means to determine the scaling factor of Eq.~(\ref{eq:def_rhl}). 
 The experimental equivalent of Eq.~(\ref{eq:def_rhl}) can be expressed as:
\begin{equation}\label{eq:double_ratio_rhl}
   \begin{split}
    \Rhl =  \frac{(\rd\xsB/\rd\pt)|_{\text{high}}}{(\rd\xsB/\rd\pt)|_{\text{low}}}\bigg/\frac{(\rd\xsZ/\rd\pt)|_{\text{high}}}{(\rd\xsZ/\rd\pt)|_{\text{low}}} .
  \end{split}
\end{equation}
 with \CSS for the \PZ bosons measured from the same \pPb data set.
 The \PZ boson reconstruction, selection requirements, and signal extraction are detailed in Ref.~\cite{CMS:2021ynu}, while cross sections results are shown in \refsupplemental.
 The measurement of \PBp nuclear modification \Rhl is shown in the \cmsLeft panel of Fig.~\ref{fig:z2mm_ratios_scaled_gf_GF} as a function of \pt, 
 while the \cmsRight panel presents it as a  function of the multiplicity density for the full range of the \PBp meson transverse momentum.
 The sources of systematic uncertainties for \Rhl are the same as those for the single cross sections. 
 The uncertainties in the double ratio are calculated by considering the correlations between the differential cross sections computed in different multiplicity bins.
 The \Rhl results as a function of \pt and the charged-particle density are consistent with unity at the current level of precision,
 providing new constraints on possible presence of medium effects in the collisions studied.
 To test the sensitivity of the \Rhl result in the highest multiplicity bin (110--250), 
 this bin was split into two parts with similar numbers of events. 
 The new highest multiplicity bin, 140--250, yields $\Rhl = 1.045\pm
 0.066\stat\pm 0.037\syst$ at an average multiplicity of 174. 
 This result is also consistent with unity but with statistical uncertainty 26\% larger than the one in the measurement using the multiplicity bin 110--250.
 The approach proposed in this Letter can be used to study medium effects in other types of collisions, and particularly  
 in systems, where unbiased determination of binary scaling by other means is unreliable.

In summary, the \PBp meson differential production cross sections have been measured as a function of the meson transverse momentum
 in proton-lead collisions at a nucleon-nucleon center-of-mass energy of 8.16\TeV.
The exclusive $\BuJpsiK{}$ decay channel was used to reconstruct the \PBp mesons.
 The fixed-order next-to-leading logarithmic predictions are in agreement with the inclusive measurement.  
 The \PBp cross section has been measured for the first time in different classes of charged-particle multiplicity in \pPb collisions,
revealing an expected rising trend in the \PBp production with increasing charged-particle multiplicity.

The scaled ratio of \PBp meson cross sections in high- to low-multiplicity events were used to study nuclear medium effects on heavy flavor production.
The scaling factors, characterizing the relative number of binary collisions between event classes of different multiplicities, were obtained,
for the first time, in a data-driven way from ratios of the measured \PZ boson cross sections in the same events.
 The observed ratios are consistent with unity within uncertainties for all event selections studied, 
 setting constraints on possible presence of medium effects in the \PBp meson production. 
 The novel approach relying on the \PZ boson measurements to extract the scaling factors provides future opportunities for medium effect searches, particularly in small collision systems.

\begin{acknowledgments}
We congratulate our colleagues in the CERN accelerator departments for the excellent performance of the LHC and thank the technical and administrative staffs at CERN and at other CMS institutes for their contributions to the success of the CMS effort. In addition, we gratefully acknowledge the computing centers and personnel of the Worldwide LHC Computing Grid and other centers for delivering so effectively the computing infrastructure essential to our analyses. Finally, we acknowledge the enduring support for the construction and operation of the LHC, the CMS detector, and the supporting computing infrastructure provided by the following funding agencies: SC (Armenia), BMBWF and FWF (Austria); FNRS and FWO (Belgium); CNPq, CAPES, FAPERJ, FAPERGS, and FAPESP (Brazil); MES and BNSF (Bulgaria); CERN; CAS, MoST, and NSFC (China); MINCIENCIAS (Colombia); MSES and CSF (Croatia); RIF (Cyprus); SENESCYT (Ecuador); ERC PRG, RVTT3 and MoER TK202 (Estonia); Academy of Finland, MEC, and HIP (Finland); CEA and CNRS/IN2P3 (France); SRNSF (Georgia); BMBF, DFG, and HGF (Germany); GSRI (Greece); NKFIH (Hungary); DAE and DST (India); IPM (Iran); SFI (Ireland); INFN (Italy); MSIP and NRF (Republic of Korea); MES (Latvia); LMTLT (Lithuania); MOE and UM (Malaysia); BUAP, CINVESTAV, CONACYT, LNS, SEP, and UASLP-FAI (Mexico); MOS (Montenegro); MBIE (New Zealand); PAEC (Pakistan); MES and NSC (Poland); FCT (Portugal); MESTD (Serbia); MCIN/AEI and PCTI (Spain); MOSTR (Sri Lanka); Swiss Funding Agencies (Switzerland); MST (Taipei); MHESI and NSTDA (Thailand); TUBITAK and TENMAK (Turkey); NASU (Ukraine); STFC (United Kingdom); DOE and NSF (USA).  
\end{acknowledgments}

\bibliography{auto_generated}
\ifthenelse{\boolean{cms@external}}{}{
\clearpage
\appendix
\numberwithin{table}{section}
\numberwithin{figure}{section}
\section{Supplemental material\label{this:supplemental}}

\subsection{Numerical differential cross section results for the \texorpdfstring{\PBp}{B+} meson in bins of \texorpdfstring{\pt}{pt}}
Table~\ref{table:Systematics_Fonll_sup} provides the numerical differential cross sections results in bins of \pt as shown in\cmsfigone{} (left)\whimA.
\begin{table}[htbp] 
\centering
	\topcaption{Summary of differential cross sections ($\mu\mathrm{b}\GeV^{-1}$) as a function of \pt(\Bu) in \pPb collisions at $\sqrtsNN=8.16\TeV$.
 The uncertainties in the integrated luminosity measurement and the \PBp meson branching fraction are not included in the systematic uncertainties.
	}
\label{table:Systematics_Fonll_sup} 
\begin{scotch}{lccc} 
\pt (\GeVns)  & $\CSS $  & Stat. Unc.        & Syst. Unc.    \\ 
\hline 
3--7 &  1423  & 157 & 76 \\
7--10 &  647  & 39 & 18 \\
10--15 &  202  & 7 & 7 \\
15--20 &  51.4  & 2.0 & 2.5 \\
20--30 &  11.3  & 0.5 & 1.0 \\
30--50 &  1.35  & 0.09 & 0.20 \\
\end{scotch} 
\end{table} 

\subsection{\texorpdfstring{$\PZ\to\PGmm\PGmp$}{Z to mu- mu+} differential cross section measurement}
The differential cross section of $\PZ\to\PGmm\PGmp$ decays is measured in \pPb collisions at a center-of-mass energy of 8.16\TeV per nucleon. 
The \PZ boson reconstruction, selection requirements, the signal extraction, and the definition of the $\PZ \to \PGmm\PGmp$ cross section are described in Ref.~\cite{CMS:2021ynu}. 
The detection efficiencies are evaluated using the same procedure used for the \PBp meson. 
The identical \pt binning, the definition of multiplicity and its classes, 
as well as the tracking efficiency utilized in the cross section measurement of the \PBp meson are also applied to the corresponding measurement with the \PZ boson.  
The same systematic sources as in the \PBp meson measurement are considered. 
The differential cross section measurement of $\PZ\to\PGmm\PGmp$ decays in \pt bins, and  for different multiplicity classes is shown in Fig.~\ref{fig:z2mm_css_Ntrk}\whimB.  
\begin{figure}[hbtp!]
  \centering
  \includegraphics[width=0.49\textwidth]{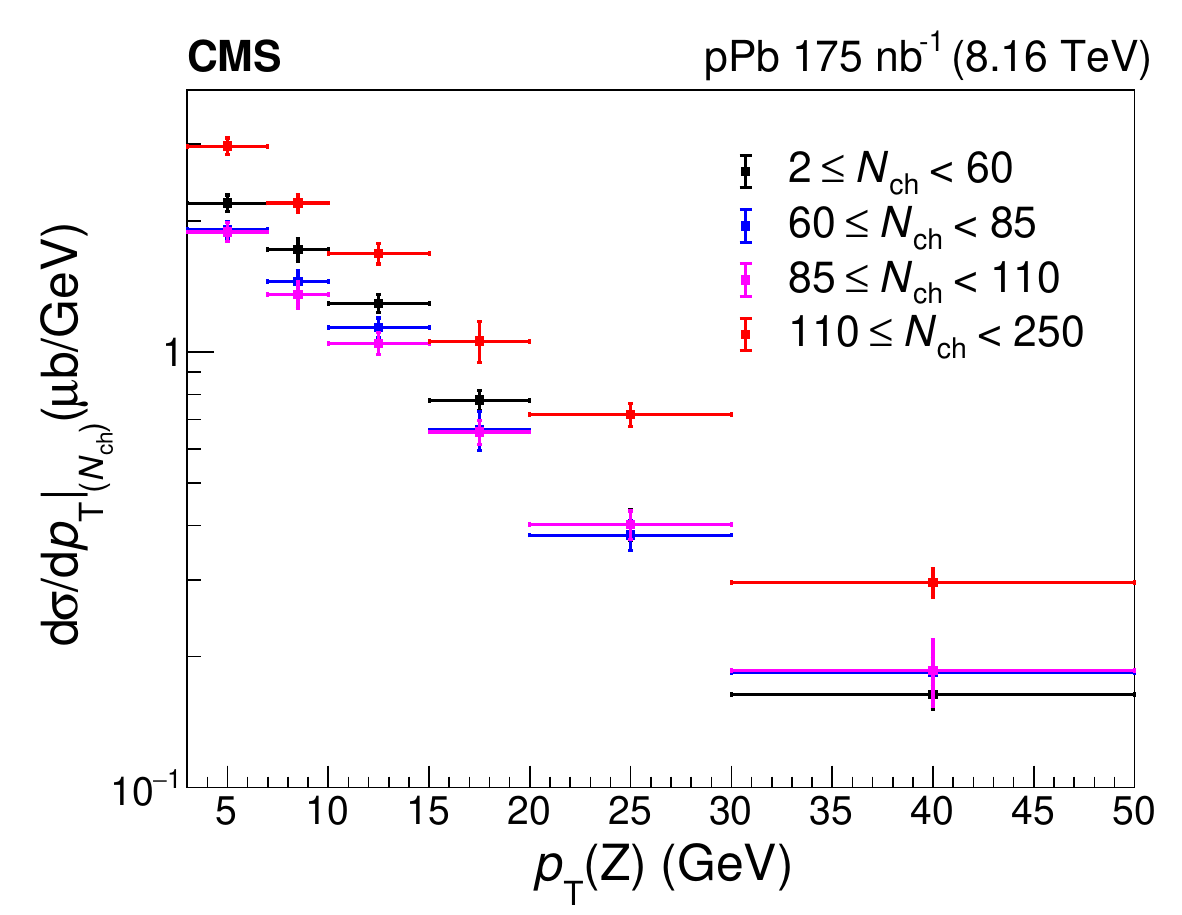}
  \caption{\PZ boson differential cross section in \pt bins for the different multiplicity classes. 
	The vertical bars represent the total uncertainties. 
  }
  \label{fig:z2mm_css_Ntrk}
\end{figure} 

}

\cleardoublepage \section{The CMS Collaboration \label{app:collab}}\begin{sloppypar}\hyphenpenalty=5000\widowpenalty=500\clubpenalty=5000
\cmsinstitute{Yerevan Physics Institute, Yerevan, Armenia}
{\tolerance=6000
A.~Hayrapetyan, A.~Tumasyan\cmsAuthorMark{1}\cmsorcid{0009-0000-0684-6742}
\par}
\cmsinstitute{Institut f\"{u}r Hochenergiephysik, Vienna, Austria}
{\tolerance=6000
W.~Adam\cmsorcid{0000-0001-9099-4341}, J.W.~Andrejkovic, T.~Bergauer\cmsorcid{0000-0002-5786-0293}, S.~Chatterjee\cmsorcid{0000-0003-2660-0349}, K.~Damanakis\cmsorcid{0000-0001-5389-2872}, M.~Dragicevic\cmsorcid{0000-0003-1967-6783}, P.S.~Hussain\cmsorcid{0000-0002-4825-5278}, M.~Jeitler\cmsAuthorMark{2}\cmsorcid{0000-0002-5141-9560}, N.~Krammer\cmsorcid{0000-0002-0548-0985}, A.~Li\cmsorcid{0000-0002-4547-116X}, D.~Liko\cmsorcid{0000-0002-3380-473X}, I.~Mikulec\cmsorcid{0000-0003-0385-2746}, J.~Schieck\cmsAuthorMark{2}\cmsorcid{0000-0002-1058-8093}, R.~Sch\"{o}fbeck\cmsorcid{0000-0002-2332-8784}, D.~Schwarz\cmsorcid{0000-0002-3821-7331}, M.~Sonawane\cmsorcid{0000-0003-0510-7010}, S.~Templ\cmsorcid{0000-0003-3137-5692}, W.~Waltenberger\cmsorcid{0000-0002-6215-7228}, C.-E.~Wulz\cmsAuthorMark{2}\cmsorcid{0000-0001-9226-5812}
% ...
\par}
\vskip\cmsinstskip
\dag:~Deceased\\
$^{1}$Also at Yerevan State University, Yerevan, Armenia\\
$^{2}$Also at TU Wien, Vienna, Austria\\
% ...
\end{sloppypar}
\end{document}